\documentclass[aps,prb,showpacs,twocolumn,preprintnumbers,amsmath,amssymb]{revtex4-1}

\usepackage{graphicx}
\usepackage{dcolumn}
\usepackage{bm}
\usepackage{hyperref}
\usepackage[mathlines]{lineno}

\usepackage{color, epstopdf, float}

\newcommand{\ket}[1]{|#1\rangle}

\newcommand{\av}[1]{\langle#1\rangle}

\newcommand{\sech}[1]{\textrm{sech}\left(#1\right)}
\newcommand{\eq}[2]{\begin{equation} \label{#1} #2 \end{equation}}

\newcommand{\kk}{\mathbf{k}}
\newcommand{\pp}{\mathbf{p}}

\newcommand{\JJ}{\mathbf{J}}

\newcommand{\AAA}{\mathbf{A}}

\newcommand{\qt}[1]{{#1^{\xi}_{\lambda}}_{\kk}(t)}

\newcommand{\kkett}[1]{\ket{{{#1}^{\xi}_{\lambda}}_{\kk}(t)}}
\newcommand{\nin}{\noindent}

\begin{document}

\title{The Nonlinear Optical Effects of Opening a Gap in Graphene}

\author{David N. Carvalho}
\author{Fabio Biancalana}
\affiliation{School of Engineering and Physical Sciences, Heriot-Watt University, EH14 4AS Edinburgh, UK \\}

\author{Andrea Marini}
\affiliation{
 ICFO-Institut de Ciencies Fotoniques, The Barcelona Institute of Science and Technology, 08860 Castelldefels (Barcelona), Spain} 

\date{\today}

\begin{abstract}
Graphene possesses remarkable electronic, optical and mechanical properties that have taken the research of two-dimensional relativistic condensed matter systems to  prolific levels. However, the understanding of how its nonlinear optical properties are affected by  relativistic-like effects has been broadly uncharted. It has been recently shown that highly-nontrivial currents can be generated in free-standing samples, notably leading to the generation of even harmonics. Since graphene monolayers are centrosymmetric media, for which such harmonic generation is deemed inaccessible, this light-driven phenomenon is both startling and promising. More realistically, graphene samples are often deposited on a dielectric substrate, leading to additional intricate interactions. Here, we present a treatment to study this instance by gapping the spectrum and we show this  leads to the appearance of a Berry phase in the carrier dynamics. We analyse the role of such a phase in the generated nonlinear current and conclude that it suppresses odd-harmonic generation. The pump energy can be tuned to the energy gap to yield interference among odd harmonics mediated by interband transitions, allowing even harmonics to be generated. Our results and general methodology pave the way for understanding the role of gap-opening  physical factors in the nonlinear optics of hexagonal two-dimensional lattices. 
\end{abstract}

\maketitle


\section{\label{sec:introduction}Introduction}

The physics of graphene is unusual in that its electrons can be adequately modelled as relativistic massless Dirac fermions, which admit a linear energy dispersion  -- the famous Dirac cones. This property itself is known to induce highly nonlinear dynamics for light \cite{Mikhailov2007}. Since this electronic dispersion is ungapped, with the bands extrema touching at the Dirac points (termed $\mathbf{K}$ and $\mathbf{K'}$), graphene behaves like a zero-gap semiconductor. \\
However, this property is only expected for free-standing, pristine graphene samples. More physically realisable samples are normally deposited on particular dielectric substrates. These intrinsic factors are known to modify the electronic and optical properties of the sample and can be successfully taken into account by simply opening a gap in the two-band spectrum \cite{Zhou2007}. 
Using various synthesis and preparation techniques, impurities, local lattice defects and vacancies \cite{Kang2008}, and strain effects \cite{Ni2008} may be physically realised and have also been shown to gap the spectrum. More challenging procedures to achieve this rely on electric biasing of graphene bilayers \cite{Castro2007} and monolayer nanostructuring into nanoribbons \cite{Li2008}. The appearance of a gap can also be conceptualised with a staggered sublattice potential, in which each triangular sublattice of the honeycomb lattice admits opposite non-zero on-site potentials (for instance when graphene is deposited on hexagonal boron nitride). \\

Each process admits a characteristic gap scale. Substrate-induced effects seem to be the most efficient to open a gap which, with the aid of ARPES measurement techniques, has been estimated to be $0.26$ eV for epitaxially-grown graphene on silicon carbide (SiC) \cite{Zhou2007}. 
Density Functional Theory calculations estimate monolayer graphene can acquire a gap of $0.35$ eV when deposited on a SiO$_2$ substrate \cite{Shemella2009}. 
Note that the extent of such a gap opening is linked to the relative geometrical configurations of the substrate and the sample alongside the dominant chemical bonds in their interaction. For instance, graphene deposited on Si-terminated silica surface with inactive
dangling bonds has been proposed as a configuration to retrieve the linear, gapless dispersion typical of free-stranding graphene \cite{Kang2008bis}. 
The transition to a semiconducting regime leads to substantially different optoelectronic features for which devices such as graphene-based transistors and photodetectors rely on \cite{Schwierz2010}. The optical behaviour of the plane-confined carriers is further modified by excitonic effects, in turn caused by screening mechanisms. These may be appreciated through theoretical models of the optical conductivity spectra and phenomenological dependence on disorder and imperfections in Ref. \cite{Pedersen2009}. 

Although advancing, the theoretical understanding of these effects on the \emph{ultrafast nonlinear} optical properties of graphene remain broadly uncharted. In this paper, we investigate the role of the energy gap in the ultrafast generation of high harmonic radiation along with related nonlinear processes, within a semiclassical quasirelativistic formalism, by explicitly solving the Dirac equation modelling the carrier dynamics. 

As a centrosymmetric material, graphene should not allow the generation of even harmonics. However, intense and ultrashort pulses provide a regime where odd harmonics interfere generating even harmonics, once gapped. Besides, the dynamical centrosymmetry breaking mechanism, a field-driven effect that globally displaces the electronic dispersion and breaks the centrosymmetry  $\mathbf{k} \leftrightarrow - \mathbf{k}$ (and consequently $\mathbf{r} \leftrightarrow - \mathbf{r}$ in direct space), has been shown to predict second-harmonic generation in freestanding, ungapped graphene \cite{Carvalho2016}.
Gapping the spectrum renders the electrons massive and, as will be demonstrated, induces a Berry phase in the carrier dynamics. Such a phase leads to interesting qualitatively different optical behaviour which is studied by extending the Dirac-Bloch equations and their framework, previously applied to massless Dirac fermions (namely monolayer graphene) \cite{Ishikawa2013, Ishikawa2010, Carvalho2016}, to incorporate a gapped Dirac spectrum.

\section{\label{sec:quasirelativisticdynamics}Quasirelativitic Dynamics}

\begin{figure}
\includegraphics[width=8cm]{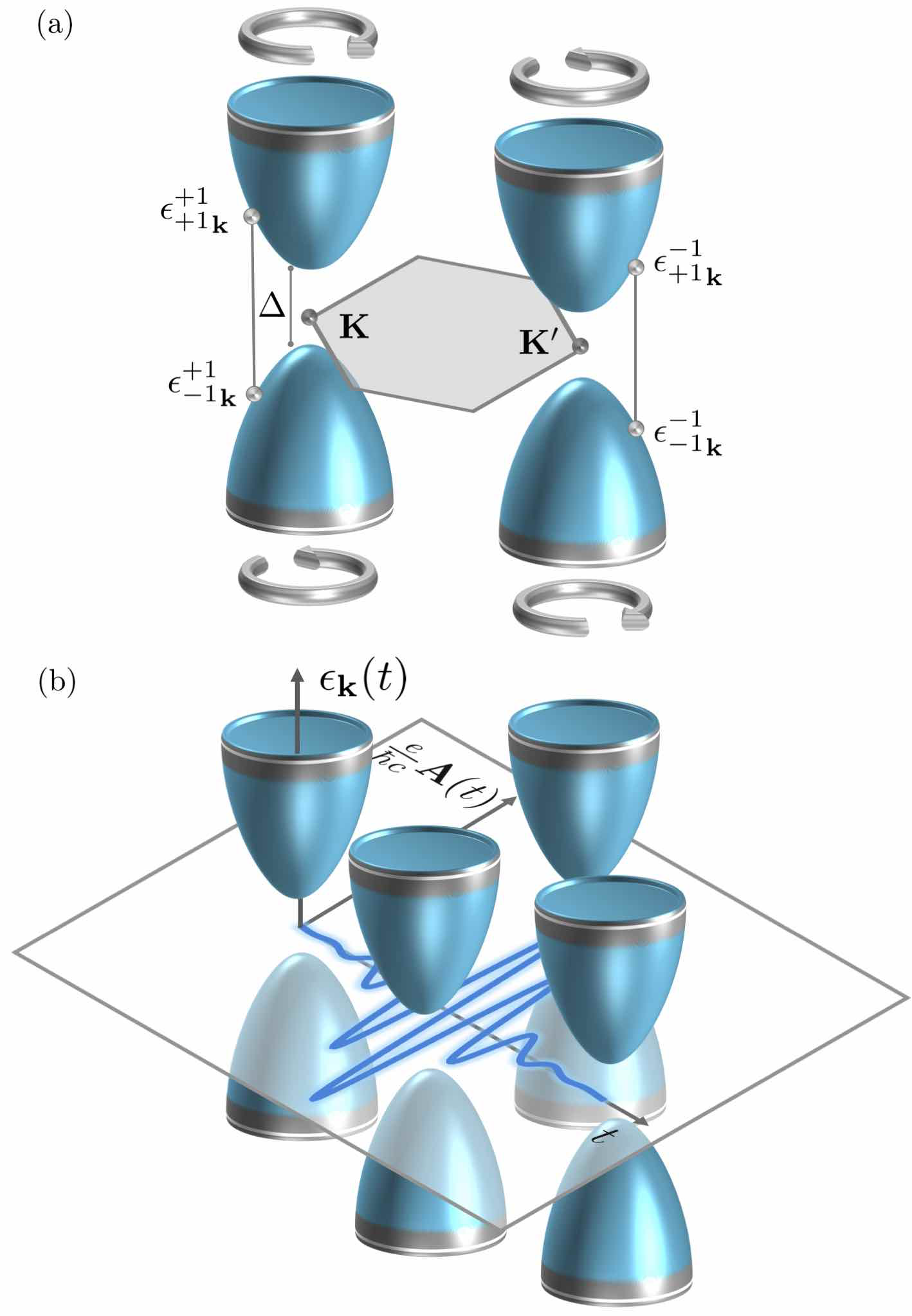}
\caption{\label{fig1}(colour online) (a) Sketch of the Hamiltonian spectra for both valleys in the low-momentum regime. Each valley admits two bands, gapped by $\Delta$. The relative sign of the field-induced Berry phase is represented by the silver arrows. (b) Depiction of the time-dependent electronic dispersion in momentum space, as given in Eq.~(\ref{dispersion}), for a particular valley. Note that the pulse shifts the dispersion globally by the time-dependent photon momentum $\mathbf{A}(t)$. This field-driven effect is only appreciable for ultrashort and intense pulses.} 
\end{figure}

Graphene is a two-dimensional crystal composed of carbon atoms and disposed in a honeycomb lattice. This arrangement stems from particular orbital hybridisation and strong covalent in-plane bonding. Linearisation of the electronic dispersion computed from tight-binding methods yields a linear dependence, which vanishes at two non-equivalent Dirac points in momentum space termed $\mathbf{K}$ and $\mathbf{K'}$. Such a linear dispersion admits a conduction and valence bands which are symmetric and touch at the Dirac points, rendering the monolayer a zero-gap system. However, a gap can be opened at the Dirac points, which are located on the edge of the Brillouin zone and shown in Fig.~\ref{fig1}(a). For low-momentum states around these points, the dispersion attains its extrema (such regions are called valleys) and carriers can be endowed with an effective mass. 

Unless an imbalance is physically realised, for instance through an electric field bias or sample inhomogeneities, intervalley scattering is highly unlikely\cite{GrapheneCarbonTwoDimensions}, as it requires exceedingly large phonon momenta, roughly of the order of the separation $|\mathbf{K}-\mathbf{K'}|$ . It can thus be reasonably assumed that the dynamics of both valleys is decoupled of each other and the carriers can be endowed with an additional degree of freedom, the \emph{valley isospin} $\xi$, where $\xi = +1$ $(-1)$ refers to states in the $\mathbf{K} \ (\mathbf{K'})$ valley. A further degree of freedom, the \emph{pseudospin} $\lambda$, with $\lambda = +1$ $(-1)$ denoting conduction (valence) band states, distinguishes between electron and hole states. \\
\nin In order to understand light-matter interactions in this gapped structure, we proceed by obtaining the wavefunction of an electron of effective mass $m \equiv \Delta/(2 v_{\rm F}^2)$ and momentum $\pp = \hbar \kk$ in the vicinity of a particular Dirac point in valley $\xi$, which must obey a two-dimensional Dirac equation:
\eq{dirac}{
i \hbar \partial_{t} \ket{\Psi^{
\xi}_{\kk}(t)} = H^{\xi}_{\kk}(t)  \ket{\Psi^{\xi}_{\kk}(t)}.}
$\Delta$ is the energy dispersion gap and $v_{\rm F} \approx c/300$ the electronic Fermi velocity. To obtain the appropriate Hamiltonian for such interactions, the canonical momentum is introduced through the minimal substitution $\pp \mapsto \pp + (e/c) \AAA(t) \equiv \bm{\pi}_{\kk}(t)$ in the field-free Hamiltonian.  yielding: \begin{equation}
\label{hamiltonian}
H^{\xi}_{\mathbf{k}}(t) = v_{\rm F}~\left(\bm{\sigma}(\xi) \cdot \left(\pp + \frac{e}{c} \AAA(t) \right) \right) + \frac{\Delta}{2} \sigma_{z},\end{equation}
where $\bm{\sigma}(\xi) \equiv (\xi \sigma_{x},  \sigma_{y})$ is a vector comprised of the 2D Pauli matrices, $e > 0$ is the absolute value of the electron charge and $c$ is the speed of light in vacuum. \\
\nin The pulse is further assumed to be normally-incident and linearly polarised along an arbitrary direction, here taken along $\hat{x}$. Its electromagnetic vector potential $\AAA$, which is chosen to satisfy the Coulomb gauge $\nabla \cdot \AAA = 0$, can thus be written as $\AAA(t) = (A(t),0,0)$. Consequently, the canonical momentum becomes $\bm{\pi}_{\kk}(t) = (p_{x} + (e/c)A(t), p_y)$.  \\

\nin Inconveniently, general analytical solutions of Eq.~(\ref{dirac}) cannot be obtained due to the time dependence of the Hamiltonian through the external parameter $\AAA(t)$. To tackle this, a general \emph{ansatz} is constructed through expansion over a basis comprised of the so-called \emph{instantaneous eigenstates} -- two linearly-independent spinors which satisfy $H^{\xi}_{\kk}(t)\kkett{u} = \qt{\epsilon} \kkett{u}$. If orthonormalised i.e. $\av{{u^{\xi}_{\lambda}}_{\kk}(t)|{u^{\xi}_{\lambda'}}_{\kk}(t)} = \delta_{\lambda \lambda'}$, they take the form:
\eq{kets}{
\kkett{u} = \frac{v_{\rm F} |\bm{\pi}_{\kk}|}{\sqrt{ \epsilon_{\mathbf{k}} (\lambda \Delta + 2 \epsilon_{\mathbf{k}})}} \begin{pmatrix}
\left(\frac{\lambda \Delta + 2 \epsilon_{\mathbf{k}}}{2 \xi v_{\rm F}|\bm{\pi}_{\kk}|} \right)e^{-i \xi \theta_{\mathbf{k}}/2} \\ 
\lambda e^{i \xi \theta_{\kk}/2}
\end{pmatrix},}
where $ \theta_{\kk}(t) = \arctan(p_{y}/[p_{x} + \frac{e}{c}A(t)])$ is the \emph{dynamical angle} of the canonical momentum vector and $\epsilon_{\kk}(t)$ the positive branch of their \emph{instantaneous} energy: 
\eq{dispersion}{ \qt{\epsilon} = \lambda \sqrt{ \left(\frac{\Delta}{2}\right)^2 + ( v_{\rm F} |\bm{\pi}_{\kk}(t)|)^2 } \equiv \lambda \epsilon_{\mathbf{k}}(t).}
\nin As seen in Fig.~\ref{fig1}(a), the spectra of both valley Hamiltonians are \emph{globally} equivalent. These solutions have a straightforward interpretation: for a particular valley $\xi$, electron and hole states exist  respectively in the conduction ($\lambda = +1$) and valence ($\lambda = -1$) bands, which are gapped by $\Delta$. The upper and lower components of the spinor can be construed as amplitudes in each of the triangular sublattices that decompose the honeycomb lattice. \\
\nin The addition of the gap leads to an inequivalence of these sublattices and consequently the appearance of a Berry phase, whose derivative can be obtained through the instantaneous eigenstates of Eq.~(\ref{kets}) as $ \dot{\gamma}^{\xi}_{\lambda \kk}(t) = i \av{\qt{u}|\partial_{t}|\qt{u}} = \xi \lambda \left( \Delta \dot{\theta}_{\kk}(t)/(4 \epsilon_{\kk}(t)) \right) \equiv \xi \lambda \dot{\gamma}_{\kk}(t)$.
The associated wavefunctions of these four states, solutions of the Dirac equation, must evolve in time as:
\eq{bandwavefunction}{
\kkett{\psi} = \kkett{u}e^{-i \lambda \Omega_{\kk}(t)}e^{i \xi \lambda \gamma_{\kk}(t)}}
and are therefore further phase-shifted by the \emph{dynamical} phase $\Omega_{\kk}(t) =(1/\hbar) \int_{-\infty}^{t} \epsilon_{\kk}(t')dt'$ and the \emph{Berry} phase $\gamma_{\kk}(t) = \int_{-\infty}^{t}\dot{\gamma}_{\kk}(t')dt'$, the latter taking the analytical form $\gamma_{\kk}(t) = \left(\Lambda_{\kk}(t) - \Lambda_{\kk}(-\infty) \right)/4$, with:
\begin{equation}
\label{Berry}
\Lambda_{\kk}(t) \equiv \arctan \left[ \frac{4  \Delta \epsilon_{\mathbf{k}}(t) \tan \theta_{\mathbf{k}}(t)}{ \Delta^2 - 4 \epsilon^2_{\mathbf{k}}(t) \tan^2 \theta_{\mathbf{k}}(t) } \right],
\end{equation}
where the time-independent term $\Lambda_{\kk}(-\infty) \equiv \gamma^{0}_{\kk}$ is a global phase factor arbitrarily included to ensure the Berry phase vanishes in the absence of field interaction. The time evolution of the Berry phase for particular states may be appreciated in Fig.~\ref{fig4}.
Finally, the {\em ansatz} $\ket{\Psi^{\xi}_{\kk}(t)}$ of Eq.~(\ref{dirac}) is taken through expansion over the band wavefunctions at that valley, i.e. 
\begin{equation}
\label{ansatz}
{\ket{\Psi^{\xi}_{\kk}(t)} = c^{\xi}_{+1}(t)\ket{\psi^{\xi}_{+1, \kk}(t)}} + c^{\xi}_{-1}(t)\ket{\psi^{\xi}_{-1, \kk}(t)}.
\end{equation} 
A word of caution is in order: although the gapped field-free Hamiltonian of Eq.~(\ref{hamiltonian}) (i.e. with $A(t)=0$)  arises from the breaking of the sublattice inversion symmetry, such Hamiltonian is only a first-order $\kk \cdot \pp$ approximation of the full tight-binding Hamiltonian and accounts only for its centrosymmetric part. Several works, e.g. in Refs. \cite{Xiao2012,Yao2014}, use our Hamiltonian to model transition metal dichalcogenide (TMD) monolayers, promising two-dimensional relativistic-like semiconductors lacking an inversion centre and hence non-centrosymmetric. Such an approximation for TMDs is adequate only to describe the linear optical properties of such media that are accounted by low momentum states, where this approximation is accurate. It is nonetheless clearly insufficient to accurately capture nonlinear light-matter phenomena of non-centrosymmetric two-dimensional media, for which higher-order terms in the $\kk \cdot \pp$ expansion explicitly break the centrosymmetry  $\kk \leftrightarrow -\kk$ (rendering the conduction and valence bands of Fig.~\ref{fig1}(a) asymmetric).

\section{\label{sec:massiveDBEs} The Massive Dirac-Bloch Equations}

The electron dynamics can be more easily understood by obtaining the time derivatives of $c^{\xi}_{\lambda}$ in and introducing new dynamical variables: the `population inversion' $w^{\xi}_{\kk}\equiv|c^{\xi}_{+}|^{2}-|c^{\xi}_{-}|^{2}$ and the `microscopic polarisation' $q^{\xi}_{\kk} \equiv c^{\xi}_{+} (c^{\xi}_{-})^{*} e^{-i(2 \Omega_{\kk} - \omega_{0}t)}$. The full Dirac equation (\ref{dirac}) can be recast in a more transparent set of equations, akin to the Bloch equations of a two-level system -- \emph{the Dirac-Bloch equations} (DBEs) -- which are derived and shown for the case of massless Dirac fermions in Ref. \cite{Carvalho2016}. When generalised to the massive case, they take the form:
\begin{widetext}
\begin{eqnarray}
\label{db1}
\dot{w}^{\xi}_{\mathbf{k}} + \gamma_{1} (w^{\xi}_{\kk} - {w_{\kk}}_{0}) -  \left( \frac{v_{\rm F} |\bm{\pi}_{\kk}|} {\epsilon_{\mathbf{k}}}\right) \left(2 \xi \dot{\theta}_{\mathbf{k}} \text{Im}(q^{\xi}_{\mathbf{k}} e^{i( 2 \xi \gamma_{\mathbf{k}} -  \omega_{0} t)}) + 4 \cot \theta_{\mathbf{k}} \dot{\gamma}_{\mathbf{k}}\text{Re}(q^{\xi}_{\mathbf{k}} e^{i( 2 \xi \gamma_{\mathbf{k}} -  \omega_{0} t)})\right)  = 0 \\
\label{db2}
\dot{q}^{\xi}_{\mathbf{k}} + i \left(2 \dot{\Omega}_{\mathbf{k}} - \omega_{0} -  i \gamma_{2} \right) q^{\xi}_{\mathbf{k}}  + \left( \frac{ v_{\rm F} |\bm{\pi}_{\kk}|}{\epsilon_{\mathbf{k}}} \right) \left(\cot \theta_{\mathbf{k}} \dot{\gamma}_{\mathbf{k}} + \frac{i \xi \dot{\theta}_{\mathbf{k}}}{2} \right) e^{-i( 2 \xi \gamma_{\mathbf{k}} -  \omega_{0} t)}   w^{\xi}_{\mathbf{k}}= 0.
\end{eqnarray}
\end{widetext}
Here, $\dot{\Omega}_{\kk}(t) = \epsilon_{\kk}(t)/\hbar$ and $\dot{\theta}_{\kk}(t) \equiv ep_{y}E(t)/|\bm{\pi}_{\kk}(t)|^2$. The constants $\gamma_{1 (2)} \equiv 1/T_{1 (2)}$ are phenomenological decay rates of the population inversion (microscopic polarisation) whereas ${w_{\kk}}_{0}$ is the equilibrium value of inversion; if the system is undoped, i.e. $\mu = 0$, and at temperature $T = 0$, it has ${w_{\kk}}_{0} = -1$, implying that all carriers are initially found in the valence band, regardless of their momentum. \\
Otherwise, for arbitrary doping and temperature, it becomes ${w_{\kk}}_{0} = - \sinh(y)/[\cosh(x) + \cosh(y)]$, with $y = \epsilon_{\kk}/(k_{\rm B}T)$ and $x = - \mu/(k_{\rm B}T)$.
These two newly-defined fields modelled by Eqs.~\ref{db1}-\ref{db2} depend on a particular valley but are nonetheless connected by precise relations. The real-valued inversions are equal, i.e.  $ w^{\xi}_{\mathbf{k}} = w^{-\xi}_{\mathbf{k}}$, while the complex-valued microscopic polarisations are statically shifted by the momentum-dependent phase $\gamma^{0}_{\kk}$, i.e. $ q^{\xi}_{\mathbf{k}} = e^{i \xi \gamma^{0}_{\mathbf{k}}} q^{-\xi}_{\mathbf{k}}$. In the limiting case of a vanishing gap, they satisfy $ q^{\xi}_{\mathbf{k}} = - q^{-\xi}_{\mathbf{k}}$. \\
The massive DBEs [Eqs.~(\ref{db1}-\ref{db2})] contain terms not present in their massless counterparts as derived and shown in Ref. \cite{Carvalho2016}.
In the driving term in Eq.~(\ref{db2}) (the one containing $w^{\xi}_{\kk}$), the quantity that multiplies the electric field $E(t)$ may be identified as a valley and time dependent {\em complex-valued} electric dipole moment:
\begin{equation}
\mu^{\xi}_{\kk}(t) = e v_{\rm F} \left(\frac{ \xi \sin \theta_{\kk}(t)}{2 \epsilon_{\kk}(t)} - i \frac{\Delta \cos \theta_{\mathbf{k}}(t)}{4 \epsilon^{2}_{\mathbf{k}}(t)} \right)
\end{equation}

We remark that the singularity found in the DBEs when $\theta_{\kk}(t) = 0$ is not problematic since both equations can be identically re-expressed so that no real singularities are present. \\
Finally, we emphasise that the Coulomb interactions amongst the carriers are not included in the massive DBEs. These are known to lead to Fermi velocity and energy band renormalisation \cite{QTheorySemiconductors}. Such effects can in principle be included by coupling the dynamics of two-level systems of all momenta and have been previously implemented for graphene, see for instance Ref. \cite{Malic2011}. 

\section{\label{sec:currentanalytics}Current Analytics}

Signatures of nonlinear light-matter interactions can be found and analysed through the electric current generated by the interaction between the monolayer and the pulse. Such a current admits, in general, two components: $\JJ(t) = (\JJ_x (t), \JJ_y (t))^\text{T}$. We investigate the role of the gap (and consequently the Berry phase) in the valley-dependent current contributions. To attain this, we proceed by first determining the $\mu$-component ($\mu = x,y$) of the current contribution of a particular momentum state $\pp$ in a valley $\xi$ in time domain, here termed a \emph{microscopic current} $j^{\xi}_{\mu, \kk}$, by applying the current density operator $\hat{j}^{\xi}_{\mu, \mathbf{k}}$ to the {\em ansatz} $\ket{\Psi^{\xi}_{\kk}}$ of Eq. (\ref{ansatz}):
\eq{currentsandwich}{
 j^{\xi}_{\mu, \kk} = \av{\Psi^{\xi}_{\kk}|\hat{j}^{\xi}_{\mu, \kk}|\Psi^{\xi}_{\kk}} - \av{\psi^{\xi}_{-1, \kk}|\hat{j}_{\mu, \kk}|\psi^{\xi}_{-1, \kk}}.}
Energy bands obtained with tight-binding methods must satisfy a sum rule that prevents currents in the valence bands to be produced \cite{Sipe1993}. However, since the dispersion of Eq.~(\ref{dispersion}) and spinors of Eq.~(\ref{kets}) are only applicable over a particular, low-momentum range where these are relativistic, the first current term in Eq.~(\ref{currentsandwich}) is insufficient to describe the actual current generated, as it contains unphysical divergences. The current can nonetheless be regularised through the introduction of the second term, which acts as an ad-hoc subtraction of valence band generated current. \\
By using the definition $\hat{j}^{\xi}_{\mu, \mathbf{k}} \equiv -(e/\hbar) (\partial H^{\xi}_{\kk}/ \partial k_{\mu})$, the valley-dependent current density operator is obtained for each component as $\hat{j}^{\xi}_{x, \mathbf{k}} = -(\xi e v_{\rm F}/\hbar) \sigma_{x}$ and $\hat{j}^{\xi}_{y, \mathbf{k}} = - (e v_{\rm F}/ \hbar) \sigma_{y}$. With these, the contribution of both components to the 2D microscopic current $\mathbf{j}^{\xi}_{\kk}(t) \equiv (j^{\xi}_{x, \kk}(t), j^{\xi}_{y, \kk}(t))^{\text{T}}$ as shown in Eq.~(\ref{currentsandwich}) is computed exactly as:

\begin{figure*}
\includegraphics[scale=0.43]{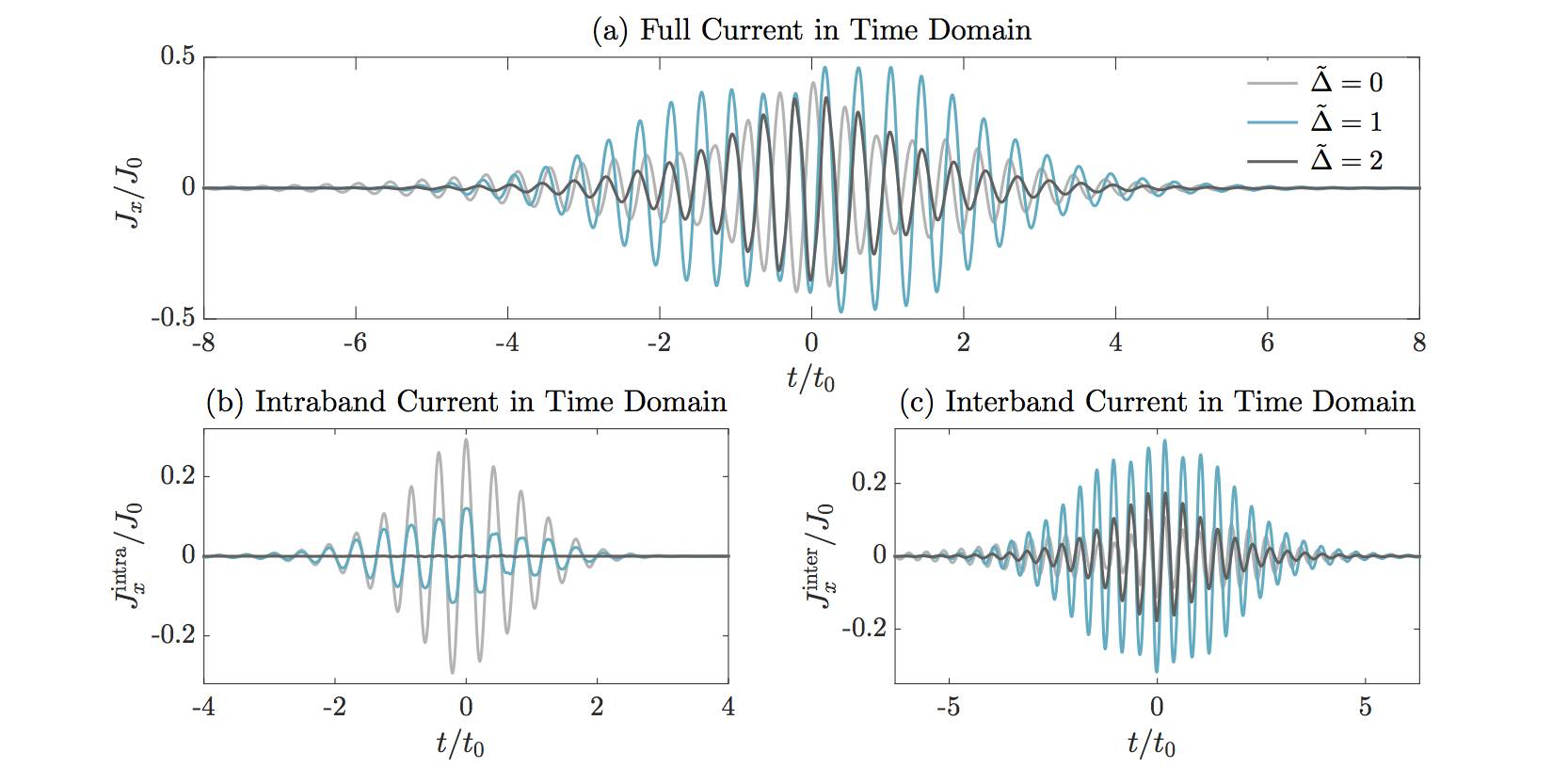}
\caption{\label{fig2}(colour online) The full photo-generated current and its separate contributions in time domain, rescaled in units of $J_{0} = -e \omega^{2}_{0}/(4 d v_{\rm F})$. (a) The total current, composed of both intraband and interband contributions. Its overall dependence on the mass stems primarily from the interband contribution. (b) The intraband current, generated from electronic transitions within the same band. Its amplitude is monotonically decreasing as the gap increases and maximal when the dispersion is ungapped. (c) The interband contribution, generated from electronic transitions across the bands. It is comprised of two terms, one being exclusively present only for gapped dispersions. The interband current amplitude is maximal when the photon energy is resonant with the gap, rapidly decreasing for larger gaps.}
\end{figure*}

\begin{figure*}
\includegraphics[scale=0.43]{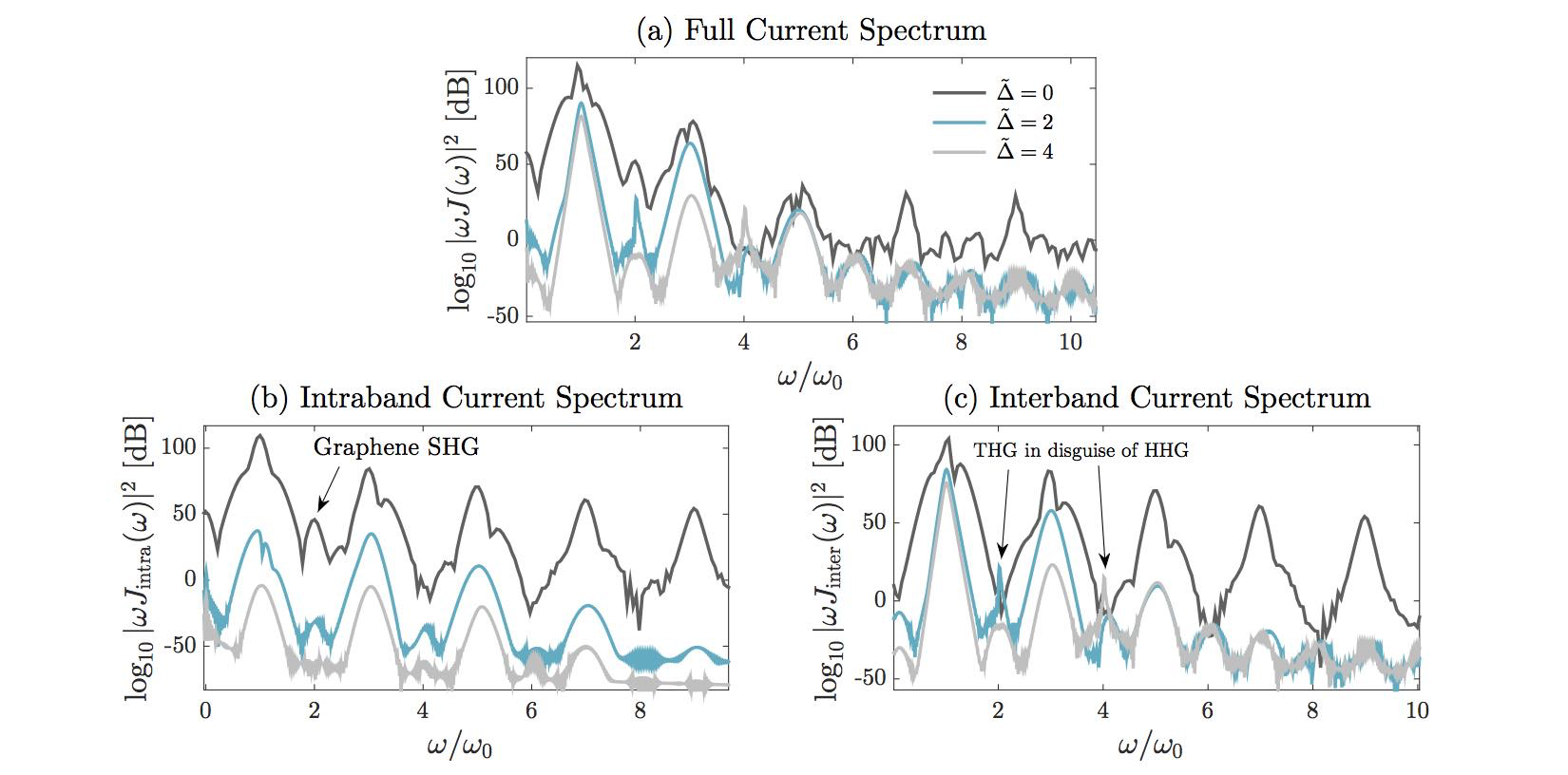}
\caption{\label{fig3}(colour online) Current spectra. (a) The total spectrum shows stronger odd harmonic generation than even harmonic generation, which can only be induced by centrosymmetry breaking mechanisms. For gapped systems, even-harmonic peaks, which are plotted in (c), are generated through third harmonic generation (THG) in disguise of higher harmonic generation (HHG). Such peaks are shown with the second and fourth-harmonic enhancements, respectively for gaps satisfying $\tilde{\Delta} = 2,4$. For vanishing gaps, even harmonic generation originates from the centrosymmetry breaking mechanism, which breaks the static centrosymmetry of the lattice and is seen in the intraband spectrum of (b). Generally, the intraband harmonic peaks decrease monotonically as the gap is increased.}
\end{figure*}

\begin{widetext}
\begin{equation}
\label{microcurrent}
\mathbf{j}^{\xi}_{\kk}(t) = -e v_{\rm F}
\begin{pmatrix}
\cos\theta_{\mathbf{k}} & \sin\theta_{\mathbf{k}} \\ 
\sin\theta_{\mathbf{k}} &  -\cos\theta_{\mathbf{k}}
\end{pmatrix}\begin{pmatrix}
\frac{v_{\rm F}|\bm{\pi}_{\kk}|}{\epsilon_{\kk}}(w^{\xi}_{\mathbf{k}} +1) - \frac{\Delta}{\epsilon_{\mathbf{k}}} \text{Re} \left(q^{\xi}_{\mathbf{k}}e^{i( 2 \xi \gamma_{\mathbf{k}} - \omega_{0}t)} \right )\\ 
-2 \xi \text{Im} \left(q^{\xi}_{\mathbf{k}}e^{i( 2 \xi \gamma_{\mathbf{k}} - \omega_{0}t)} \right )
\end{pmatrix}
\end{equation}
\end{widetext}
The physical current is finally obtained by appropriately taking all momentum contributions of both valleys into account. In the continuum limit, it is:
 \begin{equation}
\label{macrocurrent}
\JJ(t) = \frac{g_{s}}{d(2 \pi)^2} \sum_{\xi} {\int \mathbf{j}^{\xi}_{\kk}(t)}d \kk,
\end{equation}
where $d$ is the thickness of the monolayer and $g_{s}=2$ is a spin degeneracy factor, $d \kk = k dk d \phi$ is the 2-dimensional differential in momentum space and the sum is performed over both valleys.
The analytical expression in Eq.~(\ref{microcurrent}) encapsulates the \emph{exact} light-matter interactions predicted by the Dirac equation (as no approximations were applied) and displays remarkable physics richness. Two current contributions are present, depending on whether the current is originated from electronic transitions within the same band (\emph{intraband}), or across different bands (\emph{interband}).  These can be identified in Eq.~(\ref{microcurrent}) -- intraband contributions are proportional to $(w^{\xi}_{\kk} + 1)$, whereas interband contributions depend on the microscopic polarisation $q^{\xi}_{\kk}$, leading to two distinct terms. The one proportional to $(\Delta/\epsilon_{\kk})$ is a mass-induced contribution and naturally vanishes for ungapped dispersions. It can be seen that, when taking  $\Delta = 0$, both valleys contribute exactly the same to the current, i.e. $\mathbf{j}^{\xi}_{\kk}(t) = \mathbf{j}^{-\xi}_{\kk}(t)$, leading to a valley degeneracy factor $g_{\rm v} =2$ in the current of Eq.~(\ref{macrocurrent}) as previously reported in Refs. \cite{Carvalho2016, Ishikawa2013}. \\

\section{\label{sec:simulations}Simulations}

The massive Dirac-Bloch equations encapsulate a breadth of optical phenomena which become highly nontrivial in the nonlinear optical regime, once the electrons are coupled to ultrashort and intense light fields. To probe such behaviour, the graphene monolayer is pumped with a normally-incident pulse of duration $t_{0} = 31.9$ fs, central wavelength $\lambda_{0} = 4 \mu$m and frequency $\omega_{0} = 4.71 \times 10^{14}$ s$^{-1}$, photon energy $\hbar \omega_{0} = 0.31$ eV, intensity $I = 0.45$ GW/cm$^2$ and at temperature $T = 0^{\circ}$K. Additionally, realistic localised zero-averaged fields are assumed: $A(t) = A_{0} \sech{t/t_{0}} \sin{(\omega_{0}t)}$ and $E(t) = -\partial_{t}A/c$. We remark that, in order not to introduce unphysical static fields, these fields satisfy $\int_{-\infty}^{\infty}A(t)dt = \int_{-\infty}^{\infty}E(t)dt = 0 $.  \\
For an ultrashort intense pulse, the dephasing mechanisms that account for decay of the populations and polarisations (and phenomenologically accounted for by the decay rates $\gamma_{1}$ and $\gamma_{2}$) are much longer than the pulse input time $t_{0}$ and can thus be safely neglected. Time and Angle-Resolved Photoemission Spectroscopy (ARPES) techniques estimate these relaxation times as $T_1 \approx 150$ fs and $T_2 \approx 0.8$ ps. These figures are heavily affected by a combination of initial temperature, doping, pump fluence, excitation energy
and substrate type and we refer the reader to Ref.~\cite{Gierz2015} for further information on the preparation and underlying physics of the dephasing mechanisms. 
Therefore, $\gamma_{1} = \gamma_2 = 0$ are set throughout all simulations. In this coherent regime, the two-level systems in either valley are conservative, leading to a probability conservation law, namely $\partial_{t} (  4 |q^{\xi}_{\kk}|^2 + |w^{\xi}_{\kk}|^2) = 0 $, which was used to obtain numerical outputs within a strict tolerance threshold of $10^{-9}$. 

\begin{figure*}
\includegraphics[scale=0.35]{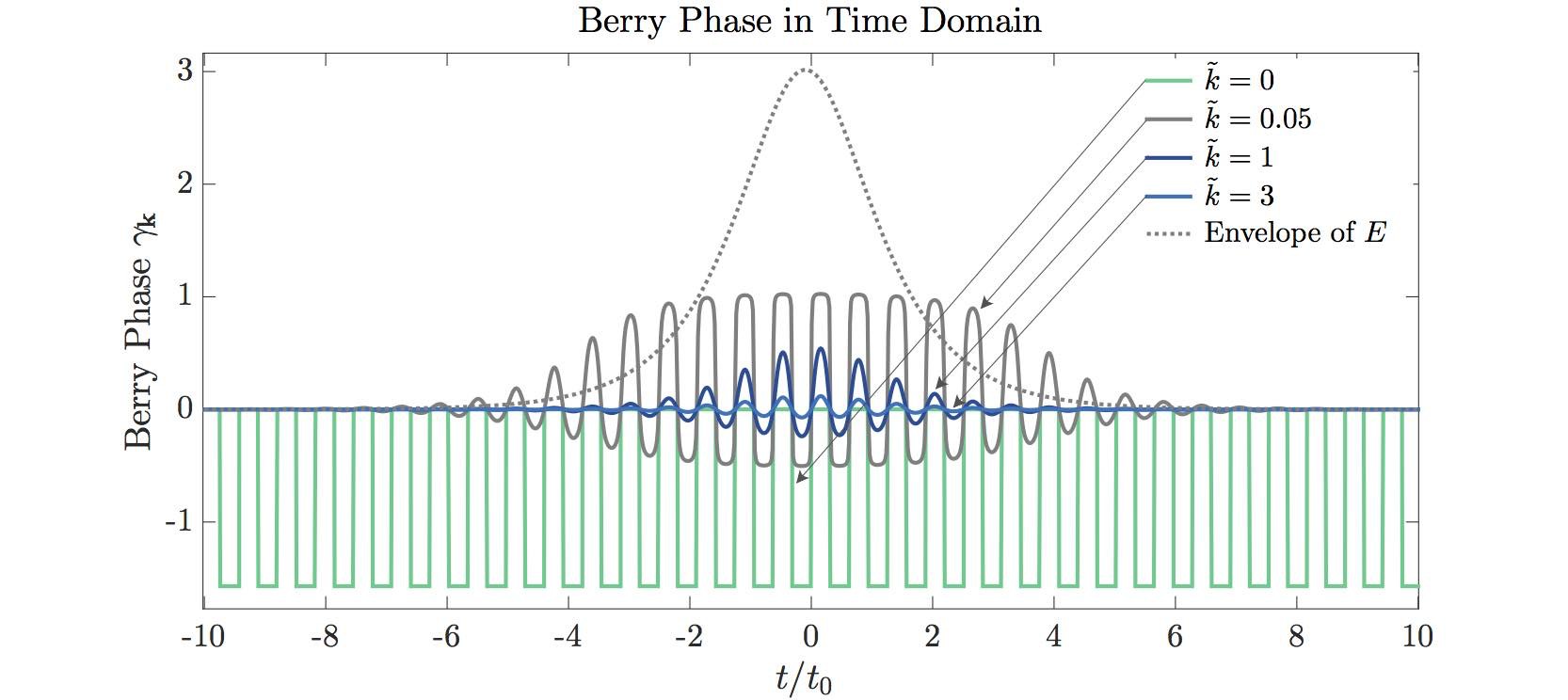}
\caption{\label{fig4}(colour online) Berry phase in time domain acquired by carriers in the conduction band in the $\mathbf{K}$ valley. With a fixed angle $\phi_{\kk} =  \pi/3$, the dependence of the phase on the rescaled momentum magnitude $\tilde{k}$ reveals highly nontrivial dynamics for momentum states close to the Dirac points, where transitions are resonant and hence strongest, showing a step-like behaviour. For off-resonant states, this phase becomes negligible as its amplitude vanishes. A juxtaposition of the electric field envelope reveals that such phase oscillations are highly asymmetrical.}
\end{figure*}

\subsection{\label{sec:currentsandspectra}Currents and respective spectra}

In order to simulate the microscopic current predicted in Eq.~(\ref{microcurrent}), the massive DBEs [Eqs.(\ref{db1}-\ref{db2})] were solved numerically with an explicit, adaptive, parallelised fourth-order Runge-Kutta algorithm.
The physical generated current is obtained once the microscopic contributions are appropriately integrated in momentum space which, when parametrised in its radial and angular components, is respectively composed of a mesh of $1000 \times 500$ states. Since this model assumes infinitely-extending bands, a radial cutoff was imposed such that all relevant microscopic contributions were accounted for. \\
The role of energy gap in the generated current is now studied with the aid of a dimensionless parameter $\tilde{\Delta} \equiv \Delta/(\hbar \omega_0)$, conveniently rescaled such that a gap satisfying $\tilde{\Delta} =1$ is exactly resonant with the pump photons. As previously mentioned, massless Dirac electrons in either valley contribute equally to the generation of current. The linearly-polarised pulse, along the $\hat{x}$ direction, does not create $\JJ_{y}$ currents which must therefore vanish identically, once their corresponding microscopic currents are integrated over all momenta and valley contributions; this is indeed observed in our simulations, and is a crucial indicator of the validity of our numerics  \cite{Carvalho2016}. In the massive regime, both components are addressed differently by the valleys, even in this simple polarisation configuration. Both valleys contribute equally to the $\JJ_{x}$ component of the current. As for the $\JJ_{y}$ component, both valleys create non-zero currents fully out-of-phase which, upon summation, cancel each other out identically. \\

The effect of the mass and Berry phase on the current may be seen in Fig. \ref{fig2}(a), where the full current in time domain $J_{x}(t)$ is shown. Its amplitude increases as the energy gap is increased, until a maximum is reached when the photon is resonant with the energy gap i.e. when $\tilde{\Delta} = 1$. Subsequently, the current amplitude vanishes for increasingly larger gaps.
 This behaviour is best understood if the intraband and interband currents are plotted separately. Fig.~\ref{fig2}(b) shows the intraband current contribution, where it can be seen that its amplitude is maximal when $\tilde{\Delta} = 0$ and monotonically decreasing with increasing energy gap. Fig.~\ref{fig2}(c) shows the interband current, itself composed of the two polarisation-dependent terms in Eq.~(\ref{microcurrent}), once integrated over momentum and valley isospin. The full current dependence on the mass stems primarily from the interband contributions, as Fig. \ref{fig2}(c) follows the pattern just described.
We remark that both interband current terms are in phase. Figs~\ref{fig2}(b,c) further reveal that the full current emerges from a very complex interplay of the competing, out-of-phase contributions of intraband and interband currents. \\
More optically pertinent information can be obtained by analysing the full current spectrum $S(\omega) = |\omega \JJ(\omega)|^2$, in dB units, versus the harmonics order $\omega / \omega_{0}$, a dimensionless parameter so that the pump pulse is centred spectrally at $\omega/\omega_{0} = 1$, which is displayed in in Fig.~\ref{fig3}(a). The spectra show strong odd harmonics being generated, commonly expected of a $\chi^{(3)}$ material. The exceedingly small peaks found for $\omega/\omega_{0} = 2, 4, ...$ on this logarithmic scale can be seen as numerical artefacts and suggest that even harmonic generation is generally absent. However, particular gap values can be seen to yield rather enhanced even-harmonic peaks. \\
In order to understand the origin of this behaviour, both the intraband and interband current spectra are respectively shown in Fig.~\ref{fig3}(b,c). For both contributions, odd-order harmonic peaks are predominant over even-order harmonic peaks, given the relatively small efficiency of even-harmonic generation expected from centrosymmetric media. The intraband current reveals second harmonic generation in ungapped samples, as previously reported in Ref. \cite{Carvalho2016}. As for the interband current, clear $n^{\text{th}}$-order harmonic peaks appear when the gap is tuned so that $\tilde{\Delta} = n$, for a positive integer $n \geq 2$. We remark that such peaks are always generated for \emph{any} gap value but will not be contribute to particular harmonic orders unless this tunability condition is met i.e. for integer $\tilde{\Delta}$. In particular, when tuned to even integers, even harmonic peaks are generated in the emission spectrum, as shown in Fig.~\ref{fig3}. Physically, the observed even harmonic peaks do not arise from $\chi^{(2)}$-like processes (occurring only in non-centrosymmetric media) but are rather understood through the coherent interference among odd harmonics, a well-known strong-field effect termed ''\emph{THG in disguise of SHG}''  occurring exclusively at the femtosecond scale. We refer the reader to pages 157-158 of Ref. \cite{Wegener2005}, as well as Ref. \cite{Tritschler2003} for more information on the physics underlying this process.

\begin{figure*}
\includegraphics[scale=0.43]{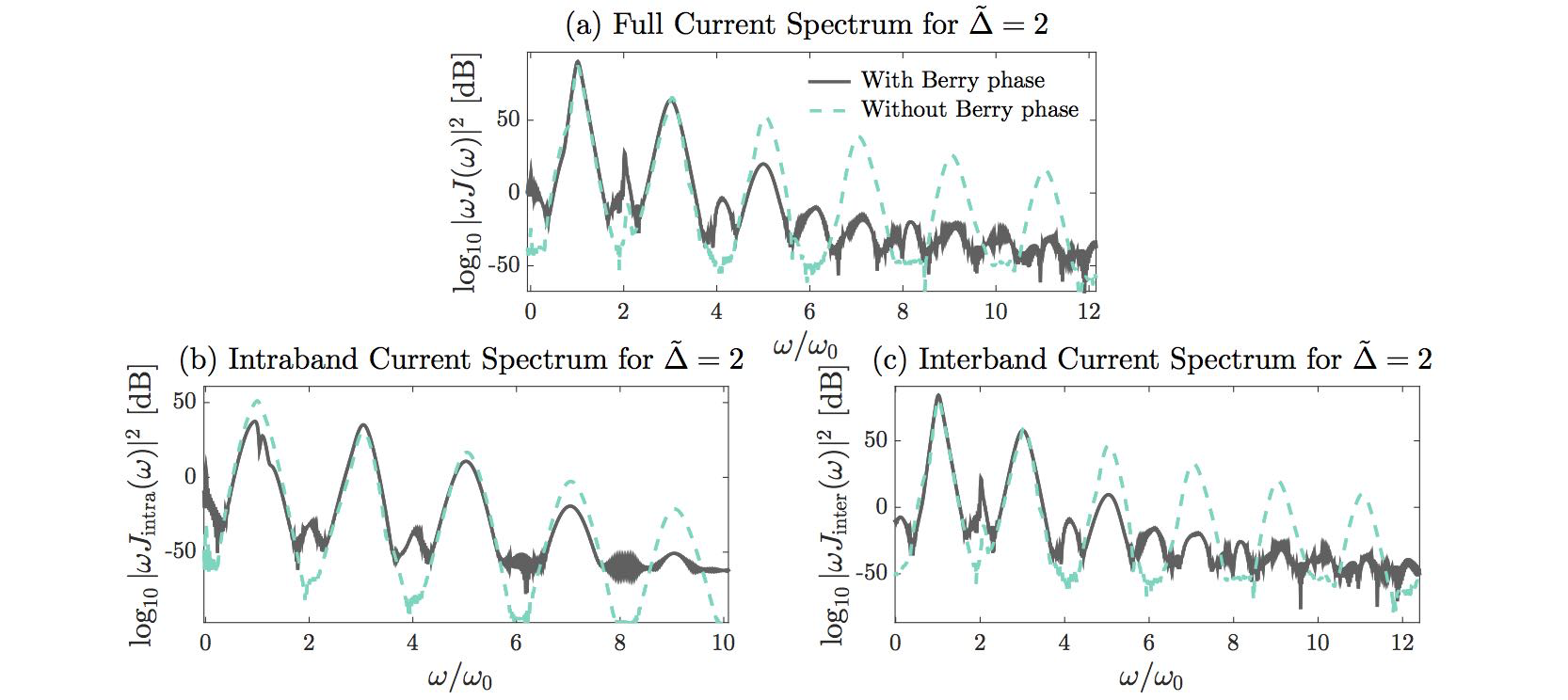}
\caption{\label{fig5}
(colour online) Comparison of current spectra in the presence or absence of the Berry phase for a gap $\Delta = 2 \hbar \omega_{0} = 0.62$ eV. The full current spectrum is shown in (a), displaying a general peak suppression of the dominant, odd harmonics. Rather negligible even harmonic peaks only exist once this phase is considered. These features are caused by the dominant, interband current, plotted in (c). The effect of the harmonic interference on the Berry phase contribution to the interband current for such a gap, here seen through the peak suppression at $\omega/\omega_{0} = 2$. The intraband spectrum, shown in (b), displays a general suppression of odd harmonics, whose extent increases as the harmonic order increases.}
\end{figure*}

\subsection{\label{sec:effectofBerryphase}The effect of the Berry Phase}

The Berry phase in Eq.~(\ref{Berry}) is only present for non-vanishing gaps and induces non-trivial contributions to the current spectra just discussed. Before engaging in determining its role in harmonic generation, the temporal dynamics of the Berry phase for various momentum states is shown in Fig.~(\ref{fig4}), by splitting the radial and angular component of their momentum vector $\kk$ respectively as $k$ as $\phi_{\kk}$, as well as rescaling $k$ to a dimensionless magnitude $\tilde{k} \equiv (2 v_{\rm F}/\omega_{0})k$. In this fashion, electrons of wavevector $\tilde{k} = 1$ have $v_{\rm F} k = \omega_{0}/2$. By juxtaposing the envelope of the electric field, it can be seen that this field-dependent phase evolves rather nontrivially, oscillates asymmetrically, and is highly dependent on which electronic state is considered. 
It is instructive to see how it changes for increasing magnitudes for a fixed angle, here arbitrarily taken as $\phi_{\kk} =  \pi/3$. High-momentum states i.e. when $\tilde{k} \geqslant 1 $, extremely detuned from the gap acquire a very small phase, which vanishes monotonically very rapidly, as the magnitude is increased. Understandably, the phase is mostly relevant for quasi-resonant states, found in the vicinity of the Dirac points i.e. $\tilde{k} = 0$, where microscopic polarisations $q_{\kk}$ are strongest. At the Dirac points, such contributions are maximal and the Berry phase undergoes continuous step-like transitions between $0$ and $-\pi/2$, independent of the field dynamics. Note that \ref{fig4} shows the phases acquired by electrons in the conduction band in the $\mathbf{K}$ valley. The relative signs acquired for each band and valley, as derived in Sec. \ref{sec:quasirelativisticdynamics}, are depicted in Fig.~\ref{fig1}(a). For instance, valence band carriers acquire a relative negative sign.  \\

The role of the Berry phase on the generation of new harmonics is now discussed. In order to achieve this, the Berry phase and its derivative are neglected by setting $\gamma_{\kk}(t) = \dot{\gamma}_{\kk}(t) = 0$ in the massive DBEs [Eqs.~(\ref{db1})-(\ref{db2})] and in the microscopic current of Eq.~(\ref{microcurrent}). This procedure is physically consistent since $\JJ_y (t)$ still vanishes after such terms are disregarded. We proceed by comparing the spectra of the full current and its intraband/interband contributions, obtained by including or excluding such terms. General features can be captured and are exemplified for a particular gap with $\tilde{\Delta} = 2$ (resulting in a realistic energy gap value of $\Delta = 0.62$ eV), whose spectra are shown in Fig.~\ref{fig5}. One can observe that the Berry phase acts on the full current , shown in Fig.~\ref{fig5}(a), and considerably suppresses odd harmonic harmonics and enhances the relevant even harmonic harmonics. The extent of the odd harmonics suppression seems to grow for higher harmonics but is much more prominent in the interband currents, where the peak differences are biggest.\\
The full behaviour can again be seen to originate from the dominant, interband contributions, plotted in Fig.~\ref{fig5}(c). The Berry phase can be identified as the agent that mostly drives odd harmonics interference (and consequently possible even harmonic generation when appropriately tuned) by considering the substantial peak enhancement at $\omega/\omega_{0} = 2$ when the phase is switched on, as previously discussed. These results can again be understood in light of the discussed THG in disguise of SHG. The intraband current spectrum comparison is shown in Fig.~\ref{fig5}(b), where it can be seen that any possibly small even harmonic peak vanishes once the Berry phase is neglected. 

\section{\label{conclusions}Conclusions} 

In conclusion, we show that relativistic two-dimensional massive fermions acquire a Berry phase when interacting with normally-incident electromagnetic pulses. This phase, whose analytical form is given and plotted for some states, only exists in the presence of an energy gap, a consequence of the inequivalence of both sublattices that decompose the honeycomb lattice of graphene. The spectrum generated by the electronic nonlinear current shows prominent odd-harmonic generation, which is generally suppressed as the energy gap is increased. Although even harmonics are generally absent for gapped dispersions, we show that their generation may be attained at the femtosecond scale through THG in disguise of SHG when the photon energy is appropriately tuned to the energy gap, generating radiation with the desired harmonic order. These processes may be conceptualised as coherent interactions of odd harmonics. Signatures of these interband-driven phenomena can be seen in the enhancement of harmonic peaks. This mechanism is to be contrasted with what has been found in pristine ungapped graphene samples, where the dynamical centrosymmetry breaking mechanism allows for even harmonics to be generated via intraband currents. We also show that the Berry phase plays a major role in the interband current dynamics and hence in the generation of even harmonics. We remark that excitonic effects are absent in the present formalism. These results and methods help establish new techniques to understand and predict the nonlinear optical behaviour of a range of two-dimensional hexagonal relativistic-like semiconductors, and help pave the way to predict quantitatively, in a generalised fashion, the effect of wide range of intrinsic or deliberate properties and phenomena, such as monolayer-substrate interactions, sample imperfections, local defects and strain effects, expected to be found in more realistic samples. \\
D.C. acknowledges insightful discussions with Dr. Lawrence Philips in Prof. Patrik Ohberg's group.

\end{document}